# Three independent one-dimensional margins for single-fraction frameless stereotactic radiosurgery brain cases using CBCT


**Qinghui Zhang**

*Department of Radiation Oncology, University of Nebraska Medical Center, Omaha, Nebraska 68198*

*Department of Medical Physics, Memorial Sloan-Kettering Cancer Center, New York, New York 10065*

**Maria F. Chan, Chandra Burman, Yulin Song**

*Department of Medical Physics, Memorial Sloan-Kettering Cancer Center, New York, New York 10065*

**Mutian Zhang**

*Department of Radiation Oncology, University of Nebraska Medical Center, Omaha, Nebraska 68198*




## ABSTRAC


**Purpose:** Setting a proper margin is crucial for not only delivering the required radiation dose to

a target volume, but also reducing the unnecessary radiation to the adjacent organs at risk. This

study investigated the independent one-dimensional symmetric and asymmetric margins between

the clinical target volume (CTV) and the planning target volume (PTV) for linac-based single-

fraction frameless stereotactic radiosurgery (SRS).

**Methods and Materials:** We assumed a Dirac delta function for the systematic error of a

specific machine and a Gaussian function for the residual setup errors. Margin formulas were

then derived in details to arrive at a suitable CTV-to-PTV margin for single-fraction frameless

SRS. Such a margin ensured that the CTV would receive the prescribed dose in 95% of the

patients. To validate our margin formalism, we retrospectively analyzed nine patients who were

previously treated with non-coplanar conformal beams. Cone-beam computed tomography

(CBCT) was used in the patient setup. The isocenter shifts between the CBCT and linac were

measured for a Varian Trilogy linear accelerator for three months. For each plan, we shifted the

isocenter of the plan in each direction by ±3 mm simultaneously to simulate the worst setup

scenario. Subsequently, the asymptotic behavior of the CTV $V_{80\%}$ for each patient was studied as

the setup error approached the CTV-PTV margin.

**Results:** We found that the proper margin for single-fraction frameless SRS cases with brain

cancer was about 3 mm for the machine investigated in this study. The isocenter shifts between

the CBCT and the linac remained almost constant over a period of three months for this specific

machine. This confirmed our assumption that the machine systematic error distribution could be

approximated as a delta function. This definition is especially relevant to a single-fraction

treatment. The prescribed dose coverage for all the patients investigated was 96.1±5.5% with an




extreme 3-mm setup error in all three directions simultaneously. It was found that the effect of the setup error on dose coverage was tumor location dependent. It mostly affected the tumors located in the posterior part of the brain, resulting in a minimum coverage of approximately72%. This was entirely due to the unique geometry of the posterior head.

40     **Conclusions:** Margin expansion formulas were derived for single-fraction frameless SRS such that the CTV would receive the prescribed dose in 95% of the patients treated for brain cancer. The margins defined in this study are machine-specific and account for nonzero mean systematic error. The margin for single-fraction SRS for a group of machines was also derived in this paper.

45     Keywords: margin, cone-beam computed tomography, image-guided radiotherapy, stereotactic radiosurgery, brain cancer

## I.   INTRODUCTION

50     Stereotactic radiosurgery (SRS) has gained increasing popularity as a treatment modality for patients with brain metastases.[1] SRS has traditionally been performed by using an invasive fixed head frame that establishes the stereotactic coordinates of the target.[2] More recently, frameless stereotactic systems have been developed and implemented with the help of an image-guided system.[3-10]

55         In a previous paper,[11] a three-dimensional (3D) margin expansion formula was proposed to derive the planning target volume (PTV) from the clinical target volume (CTV) for SRS. In



that derivation, it was assumed that the setup errors in all directions were the same. This may not be clinically realistic in most cases. However, the previous formula is valid if the maximum margin is used in all directions. This approach, nevertheless, will inevitably result in over dosing more healthy brain tissues. In this paper, we will derive the mathematical formulas for three one-dimensional (1D) expansions of the CTV–PTV margin in an effort to minimize the unnecessary dose to the organs at risk (OARs) and, in the meantime, to provide a minimally required margin for an effective and safe single-fraction treatment. A1D expansion means that the uni-dimensional margin is derived individually and applied to its corresponding principal axis of a 3D coordinate system. The differences between 1D and 3D expansions are graphically demonstrated in Fig.1. We would like to explicitly state here that the fundamental objective of this study is to derive a mathematical formula to quantitatively determine the treatment margin to account for patient setup uncertainty. The detailed computational algorithm on volume expansion is not the central point of this paper.

Reports on the CTV–PTV margins for frameless SRS using cone-beam computed tomography (CBCT) are rare. The main reason for this may be that the setup accuracy has been extensively studied and explored for image-guided radiation therapy (IGRT).[12] Thus, many people may no longer consider the CTV-PTV margin as a major clinical issue. However, the effects of the systematic isocenter shifts between the CBCT and the linac's radiation beam have not been systematically investigated. This isocenter discrepancy directly affects the exact definition of the CTV-PTV margin. This is especially true for SRS cases where the margin is normally in the order of millimeters. For multi-fraction treatment regimens, various mathematical models for margin determination have been proposed, investigated, and implemented.[13-24] However, those formulas generally vary from publication to publication.[13] For



80 example, the CTV–PTV margin recommended by the ICRU (International Commission on Radiation Units and Measurements)Report 62[14] is different from van Herk's formula.[19] A detailed discussion on the definitions of the CTV and the PTV can be found in relevant ICRU Reports[15] and also in a seminar article.[16] Owing to this, readers are kindly reminded that the determination of the CTV–PTV margin is not an exact science at this point.[16] Because the current

85 margin design still adopts an empirical approach, the physicians' clinical experience plays a crucial role in the margin determination.[16] There are many contradictions and inconsistencies among margins for frameless SRS. These are due to several reasons. One of them is that frameless SRS is a new treatment modality compared to traditional frame-based SRS. Consequently, radiation oncologists tend to apply their previous clinical practice directly to

90 frameless SRS margin design without any modification. Our recent informal survey shows that the margin for linac-based frameless SRS varies significantly, from zero to several millimeters for a dozen of academic institutions. Contrast-enhanced computed tomography (CT) and magnetic resonance imaging (MRI) can help radiation oncologists accurately delineate the tumor volume. Previous studies have found that a 1-mm margin is needed for expansion from the gross

95 tumor volume (GTV) to the CTV.[25-26] However, the technologies used in those studies are different from the one employed in this investigation. In this paper, similar to the approaches adopted by previous multi-fraction margin publications, we will ignore the delineation uncertainties and concentrate on an important, but unsolved physics problem, the effects of the systematic error between the CBCT and the linac isocenters on the margin determination.

100 Mathematically, it is not a trivial task to determine a clinically meaningful margin with many uncertainties. ICRU Report 50[15] does not recommend adding all uncertainties linearly because this would produce a much larger margin than the clinically needed. Therefore, different



margin formulas have been proposed based on different assumptions. However, all those previous studies, both theoretical and clinical ones, have been confined to multi-fraction

105    treatments only. The formulas thus derived cannot be directly applied to single-fraction SRS. In addition, because the isocenter shifts between the CBCT and the linac overlap with the setup error, one had better not simply combine the two linearly. Otherwise, the margin so constructed would be overestimated. In this paper, we will provide an alternative but novel method for margin determination, specifically for single-fraction SRS. In contrast to previously published

110    approaches, our margin formulas explicitly include the isocenter shifts as one of the systematic errors in our derivation.

We assume that frameless SRS is performed on a linac with CBCT capability. On such a premise, two coordinate systems can be established: one with the origin placed at the CBCT isocenter and the other one with the origin placed at the linac isocenter. By default, the two

115    coordinate systems use the same mathematical convention so that their corresponding axes are about parallel to each other. For IGRT-based SRS (IG-SRS), the planning CT isocenter is implicitly assumed to be at the linac isocenter. Thus, upon a successful completion of the image registration between the CBCT and the planning CT, the CBCT isocenter should, in theory, coincide with the planning CT isocenter, i.e., the linac isocenter. However, due to linac

120    mechanical limitations, CBCT isocenter shift, CT image quality degradation, and human factors, there are a number of uncertainties constantly present in the patient setup. Those uncertainties can be broadly classified into two types: (1) systematic errors, including, for example, image registration error, target delineation uncertainty, and inaccurate isocenter position; and (2) residual setup errors, including inaccurate 4-degrees-of-freedom (DOF) couch positioning (three

125    translations and one rotation) caused by a finite couch stepping precision, leading to not only



non-negligible residual setup errors, but also a preclusion of a full 6-DOF error correction (three translations and three rotations) and errors induced by an incorrect order of rotation and translation operations.

This paper is presented as follows: Section II (Methods) describes the mathematical relation between the probabilities that patients receive the prescribed dose and derives the margins for three independent 1D expansions. In addition, Section II also derives the margin formulas for a group of machines and the entire patient population. Furthermore, Section II provides the margin formulas for 3D expansion such that the CTV receives the prescribed dose in 95% of the treated patients. Section III (Results and Discussion) elaborates on the differences between our margin formulas and previously published ones and describes a procedure for measuring the isocenter shifts between the CBCT and the linac. In addition, Section III also presents our data on the CTV coverage for nine patients when the setup error approached the CTV-PTV margin simultaneously in all three directions. Section IV provides our conclusions.

## II. METHODS

### II.A. General approach

Adopting the convention widely used by previous investigators,[22-23] we make following assumptions on SRS to facilitate our margin formula derivations: (a) Variation in beam profiles within the PTV is negligible; (b) CT numbers do not change dramatically in this region; and (c) the surface curvature does not vary appreciably. Currently, our institution employs a 3D static conformal technique for single-fraction frameless SRS. The treatment dose is prescribed to the 80% isodose surface that, ideally, should cover the entire CTV, i.e., $V_{80\%} = 100\%$.



For each patient, the percent CTV covered by the prescribed dose ($V_{Dc}$) can be calculated by the following equation:

$$V_{Dc}(\vec{V}, D_c) = \frac{\int_{CTV} H(D_{CBCT}(\vec{x}) - D_c) d\vec{x}}{\int_{CTV} d\vec{x}} = \frac{\int_{CTV} H(D_{CT}(\vec{x} + \vec{V}) - D_c) d\vec{x}}{\int_{CTV} d\vec{x}} \ . \tag{1}$$

where $D_c$ is the prescribed dose, i.e., the 80% isodose surface, throughout this paper.[11] Here, the integration is performed over the points inside the CTV and $H(x)$ is a step function that equals 1 when x>0 and zero otherwise. $V_{Dc}(\vec{V}, D_c)$, representing an integral volume, is a function of $D_c$ and the CTV isocenter shift vector $\vec{V}$. $D_{CBCT}$ is the dose in the patient CBCT scan and $D_{CT}$ is the dose in the planning CT. The probability that the CTV receives $D_c$ can be expressed as[11]

$$\begin{aligned} P &= \int H(V_{Dc} - V_c) P(\vec{V}) d\vec{V} \\ &= \int H(V_{Dc} - V_c) P(\vec{V}_s, \vec{V}_r) d\vec{V}_s \ d\vec{V}_r \end{aligned} \tag{2}$$

where $H(x)$ is the aforementioned step function and $V_c$ is a threshold value of $V_{Dc}$, set as 100% in this paper. $P(\vec{V})$ is the probability distribution of the isocenter shift vector $\vec{V} = \vec{V}_s + \vec{V}_r$ for all patients and can be expressed as:

$$P(\vec{V}) = \int \delta(\vec{V} - \vec{V}_s - \vec{V}_r) P(\vec{V}_s, \vec{V}_r) d\vec{V}_s \ d\vec{V}_r \ . \tag{3}$$

Here, $P(\vec{V}_s, \vec{V}_r)$ is the probability distribution for the systematic error ($\vec{V}_s$) and residual setup error ($\vec{V}_r$).

To complete Eq. (2), we need two functions $D_{CT}(x)$ and $P(\vec{V})$. Function $D_{CT}(x)$ represents the dose that covers the CTV in the planning CT. For a 3D conformal treatment plan, the CTV coverage is its objective function, i.e., mathematically, $D_{CT}(x) \geq D_c$. The exact form of



the dose distribution function within the PTV is actually not so critical for a 3D conformal plan. Therefore, the CTV–PTV margin is determined mainly by the distribution function $P(\vec{V})$ in Eq. (2).

In a previous publication,[11] the standard deviations of the residual setup errors were assumed to be equal in all three directions ($x$, $y$, and $z$). In this section, we will provide a solution to cases that have significantly different standard deviations in these directions. Under such a circumstance, one can expand the tumor volume anisotropically, i.e., moving the PTV surface along a direction for a certain distance. Such an approach would be easily understood graphically if the tumor shape resembled a box. Given this assumption, $P(\vec{V}_s, \vec{V}_r)$ can now be explicitly written as follows:

$$P(\vec{V}_s, \vec{V}_r) = \delta(\vec{V}_s - \vec{V}_{s0}) \frac{1}{(2\pi\sigma_x^2)^{1/2}} \frac{1}{(2\pi\sigma_y^2)^{1/2}} \frac{1}{(2\pi\sigma_z^2)^{1/2}}$$
$$\times \exp\left(-\frac{(V_{rx} - V_{r0x})^2}{2\sigma_x^2}\right) \exp\left(-\frac{(V_{ry} - V_{r0y})^2}{2\sigma_y^2}\right) \exp\left(-\frac{(V_{rz} - V_{r0z})^2}{2\sigma_z^2}\right),$$

(4)

where $V_{rx}$, $V_{ry}$, and $V_{rz}$ are the three components of the residual setup errors in $x$, $y$, and $z$ directions, respectively; $\vec{V}_{s0}$ is the systematic error, which is a constant for a specific machine; $V_{r0x}$, $V_{r0y}$, and $V_{r0z}$ are the three components of the mean residual setup error ($\vec{V}_{r0}$). Here, $\vec{V}_{s0}$ represents the isocenter shifts between the CBCT and the linac. $\sigma_i (i = x, y, z)$ is the standard deviation of the setup error in the $i_{th}$ direction. The residual setup errors are included in the following cases: (1) residual setup errors with a shift correction <1 mm in each direction post-registration; (2) original setup errors that are so small that shift corrections are not needed.



**II.B. One-dimensional expansion margins for a single-fraction SRS case and a specific machine**

In this section, we will construct a model to obtain the relationship between the probability of successful treatment and the margin required for a single-fraction SRS case. In this derivation, the systematic error is assumed to be the isocenter shifts between the CBCT and the linac. The derivation can be generalized to include all those systematic errors that are constants during treatment.

Assuming unequal margins and symmetrical volume expansion in each direction, Eq. (2) can be rewritten as:

$$P = \int H(C_x - |V_x|)H(C_y - |V_y|)H(C_z - |V_z|)P(\vec{V}_s, \vec{V}_r)d\vec{V}_r d\vec{V}_s. \tag{5}$$

where $C_x, C_y,$ and $C_z$ are the margins in the $x, y, z$ directions, respectively, and $V_x, V_y,$ and $V_z$ are the three components of the isocenter shift vector $\vec{V} = \vec{V}_s + \vec{V}_r$. Defining $\vec{U} = \vec{V}_r - \vec{V}_{r0}$ and $\vec{W}_0 = \vec{V}_{s0} + \vec{V}_{r0}$, Eq. (5) becomes

$$P = \int H\left(C_x - |\vec{W}_{0x} + \vec{U}_x|\right)H\left(C_y - |\vec{W}_{0y} + \vec{U}_y|\right)H\left(C_z - |\vec{W}_{0z} + \vec{U}_z|\right)$$
$$\times \frac{1}{\left(2\pi\sigma_x^2\right)^{1/2}}\exp\left(-\frac{U_x^2}{2\sigma_x^2}\right)\frac{1}{\left(2\pi\sigma_y^2\right)^{1/2}}\exp\left(-\frac{U_y^2}{2\sigma_y^2}\right)\frac{1}{\left(2\pi\sigma_z^2\right)^{1/2}}\exp\left(-\frac{U_z^2}{2\sigma_z^2}\right)dU_x dU_y dU_z. \tag{6}$$

Eq. (6) can be rewritten as $P = P_x P_y P_z$ if we define the components $P_i (i = x, y, z)$ as follows

$$P_i = \int H\left(C_i - |W_{0i} + U_i|\right)\frac{1}{\left(2\pi\sigma_i^2\right)^{1/2}}\exp\left(-\frac{U_i^2}{2\sigma_i^2}\right)dU_i$$
$$= \frac{1}{2}\left[erf\left(\frac{C_i + W_{0i}}{\sqrt{2}\sigma_i}\right) + erf\left(\frac{C_i - W_{0i}}{\sqrt{2}\sigma_i}\right)\right] \tag{7}$$



where $erf(x) = \dfrac{2}{\sqrt{\pi}} \displaystyle\int_0^x \exp(-t^2)\, dt$ is the error function.

## II.C. One-dimensional asymmetrical expansion

Supposing that a treatment planning system has a function to allow asymmetrical

205  boundary expansion in each direction, one can define the following margin, with left-right (LR)

as the $x$ direction, anterior-posterior (AP) as the $y$ direction, and feet-head (FH) as the $z$ direction:

$$P = \int H(C_L + V_x) H(C_R - V_x) H(C_A + V_y) H(C_P - V_y)$$
$$\times H(C_F + V_z) H(C_H - V_z)\, P(\vec{V}_s, \vec{V}_r)\, d\vec{V}_s\, d\vec{V}_r. \tag{8}$$

where $C_i (i = L, R, A, P, F, \text{and } H)$ is the margin in each direction. It is easy to see that the

asymmetric margins can be converted to symmetric margins by defining $C_L = C_x - W_{0x}$,

210  $C_R = C_x + W_{0x}$, $C_A = C_y - W_{0y}$, $C_P = C_y + W_{0y}$, $C_F = C_z - W_{0z}$, and $C_H = C_z + W_{0z}$. Thus Eq.

(8) becomes:

$$P = P_x^{new} P_y^{new} P_z^{new}, \tag{9}$$

where $P_i^{new} (i = x, y, \text{and } z)$ denotes

$$P_i^{new} = erf\left(\frac{C_i}{\sqrt{2}\sigma_i}\right). \tag{10}$$

215  ## II.D. One-dimensional expansions for all machines and patients

As a comparison, we will also calculate the margins for a group of machines on which all

patients are treated with single-fraction SRS. In this scenario, the systematic errors are different



for different machines and we assume that they follow a Gaussian function, as in a majority of publications. We will determine the margins for three 1D expansions.

220 When the residual setup errors or systematic errors change dramatically in each direction, one needs to expand the representative tumor volume independently in each direction. In this case, $P(\vec{V}_s, \vec{V}_r)$ changes to

$$
\begin{aligned}
P(\vec{V}_s, \vec{V}_r) = {} & \frac{1}{\left(2\pi\,\Pi_x^2\right)^{1/2}} \frac{1}{\left(2\pi\,\Pi_y^2\right)^{1/2}} \frac{1}{\left(2\pi\,\Pi_z^2\right)^{1/2}} \exp\left(-\frac{(V_{sx}-\overline{V}_{s0x})^2}{2\Pi_x^2} - \frac{(V_{sy}-\overline{V}_{s0y})^2}{2\Pi_y^2} - \frac{(V_{sz}-\overline{V}_{s0z})^2}{2\Pi_z^2}\right) \\
& \times \frac{1}{\left(2\pi\,\overline{\sigma}_x^2\right)^{1/2}} \frac{1}{\left(2\pi\,\overline{\sigma}_y^2\right)^{1/2}} \frac{1}{\left(2\pi\,\overline{\sigma}_z^2\right)^{1/2}} \exp\left(-\frac{(V_{rx}-\overline{V}_{r0x})^2}{2\overline{\sigma}_x^2} - \frac{(V_{ry}-\overline{V}_{r0y})^2}{2\overline{\sigma}_y^2} - \frac{(V_{rz}-\overline{V}_{r0z})^2}{2\overline{\sigma}_z^2}\right).
\end{aligned}
$$

$$(11)$$

225 where $\Pi_i\,(i=x, y, \text{and } z)$ is the standard deviation of the isocenter shifts within a group of linacs. By bringing Eq. (11) into Eq. (2), we yield $P_i\ (i=x, y, \text{and } z)$ and $P = P_x\,P_y\,P_z$,

$$
\begin{aligned}
P_i &= \frac{1}{2}\left[ erf\left(\frac{C_i + \overline{V}_{s0i} + \overline{V}_{r0i}}{\sqrt{2\left(\Pi_i^2 + \overline{\sigma}_i^2\right)}}\right) + erf\left(\frac{C_i - \overline{V}_{s0i} - \overline{V}_{r0i}}{\sqrt{2\left(\Pi_i^2 + \overline{\sigma}_i^2\right)}}\right) \right] \\
&= \frac{1}{2}\left[ erf\left(\frac{C_i + \overline{W}_{0i}}{\sqrt{2\left(\Pi_i^2 + \overline{\sigma}_i^2\right)}}\right) + erf\left(\frac{C_i - \overline{W}_{0i}}{\sqrt{2\left(\Pi_i^2 + \overline{\sigma}_i^2\right)}}\right) \right].
\end{aligned}
$$

$$(12)$$

Where $\overline{W}_{0i} = \overline{V}_{s0i} + \overline{V}_{r0i}$. At this point, we have derived the relations between the margins and the probabilities that the CTV receives the prescribed dose for three 1D expansions.

230

## II.E. Margin for symmetric 1D expansion

The following procedures are used in the derivation of the margin formula for three 1D expansions: (1) For a 1D expansion, Eq. (7) is used to obtain the relation between $C_i$ and $\sigma_i$ for



a fixed $W_{0i}$ and $P_i = 0.98$. Polynomial functions are used to fit those relations and the

235  corresponding coefficients are then obtained. (2) Repeating the above process for a different $W_{0i}$,

the coefficients as a function of $W_0$ are obtained. Thus, the margin formula for a specific

machine was finally obtained. For a group of machines, due to the similarity between Eq. (12)

and Eq. (7), the derivation procedure in this case is very similar to that for a specific machine.

If $P = 0.95$, $P_x = P_y = P_z = (0.95)^{1/3} = 0.98$. We will expand the margin $C_i$ ($i$=$x$, $y$, and

240  $z$) for each direction as:

$$C_i = W_{0i} + b_1(W_{0i})\,\sigma_i + b_2(W_{0i})\,\sigma_i^2 \,. \tag{13}$$

We will first calculate $C_i$ vs. $\sigma_i$ for each different $W_{0i}$, then we will use Eq. (13) to fit the curve

to get $b_1(W_{0i})$ and $b_2(W_{0i})$. After that, we will fit $b_1(W_{0i})$ and $b_2(W_{0i})$ as a function of $W_{0i}$.

With this numerical approach, we have found that the behavior of the margin parameters could

245  be fitted in functions as depicted in Fig. 2 and Fig. 3:

$$b_1(W_{0i}) = 2.331 - 1.425\,W_{0i} + 2.296\,W_{0i}^2 - 1.539\,W_{0i}^3 + 0.374\,W_{0i}^4 \,, \tag{14}$$

and

$$b_2(W_{0i}) = 0.434\,W_{0i} - 0.917\,W_{0i}^2 + 0.676\,W_{0i}^3 - 0.171\,W_{0i}^4 \,. \tag{15}$$

Here, we need to point out that in the limit of $W_{0i} \to 0$, we obtain $b_1(0) = 2.331$ and $b_2 = 0$.

250  Therefore, the margin formula (Eq. 13) is reduced to:

$$C_{II} = 2.331\sigma \,. \tag{13a}$$

Eq. (13) is an approximated solution to Eq. (7). To confirm its accuracy, we have solved Eq. (7)

numerically for wide ranges of $\sigma$ and $W_{0i}$, from 0.0 to 1.5 mm in a step size of 0.15 mm for $W_{0i}$



and from 0.05 to 2.5 mm in a step size of 0.05 mm for $\sigma$. We have found that the greatest

255   difference is only 0.013 mm. Therefore, we can conclude with full confidence that the formula

provided in this paper can be legitimately applied to single fraction cases if the standard

deviation of the residual setup error is $\leq$ 2.5 mm and the isocenter shift is $\leq$ 1.5 mm. As a

concrete evidence, we hereby provide a plot for a typical clinical case when $W_{0i} = 0.8$ mm. The

exact and approximated solutions given by Eq. (7) and Eq. (13), respectively, are plotted in Fig.

260   4.

### II.F. Margin for Asymmetric 1D expansion

For 1D asymmetric expansion, we have

$$C_i = 2.331\,\sigma_i, \tag{16}$$

265   with $i = x$, $y$, and $z$ and the corresponding margins being $C_L = C_x - W_{0x}$; $C_R = C_x + W_{0x}$;

$C_A = C_y - W_{0y}$; $C_P = C_y + W_{0y}$; $C_F = C_z - W_{0z}$; and $C_H = C_z + W_{0z}$.

Graphically, the margin so defined is equivalent to copying the CTV to a new position

$(W_{0x}, W_{0y}, W_{0z})$ and then expanding it with a margin $C_i$. If we assume $\sigma_i = 0$, then we have a

"copy CTV". Delivering treatment using this copied CTV has significant shortcomings: the

270   relative position of this copied CTV with respect to the original patient position is not the same.

Consequently, the delivered dose distribution in the CBCT CTV will deviate from the planned

dose distribution in the original plan CTV, as demonstrated in Fig. 5. The image on the left in

Fig. 5 is the planning CT with the original CTV. To achieve the treatment planning objective, the

planning beam arrangement should cover the entire CTV adequately. Assuming that the CTV is

275   copied to or reproduced in a different location in space (the image on the right in Fig. 5) because



of the isocenter shift between the CBCT and the linac, the beams would have moved by the same amount accordingly. Therefore, during the real treatment, this isocenter discrepancy would inevitably lead to a deviated patient's dose distribution from that calculated. In theory, one could use two sets of CT scans for daily clinical cases: one for planning (the left one) and the other one for positioning (the right one). However, the current technology does not support this dual CT scan function and, in addition, this method is also impractical for a busy clinical center from a logistical point of view. This shortcoming might not be significant if $\sigma_i$ were very large. However, one should be cautious when using asymmetric expansion and should also realize that there are some uncertainties in dose calculation as well.

### II.G. Margins for all patients and all machines

The derivation of margins for all patients and all machines is quite similar to that for a specific machine. For a group of machines and the whole patient population, the margin for single-fraction SRS is

$$C_{i\,group} = \overline{W}_{0i} + b_1(\overline{W}_{0i})\sqrt{\Pi_i^2 + \overline{\sigma}_i^2} + b_2(\overline{W}_{0i})\left(\Pi_i^2 + \overline{\sigma}_i^2\right). \tag{17}$$

where $i = x$, $y$, and $z$. $b_1(\overline{W}_{0i})$ and $b_2(\overline{W}_{0i})$ have the same function format as in Eq. (14) and Eq. (15). Here, $\Pi_i$ is the standard deviation of the isocenter shifts, but it can be extended to include any systematic error. As a comparison, van Herk's formula for the 1D expansion (for 98% confidence and 98% minimal dose to the CTV) is going to be

$$C_{vanHerk} = 2.331\Sigma + 2.06\sigma. \tag{18}$$



where $\Sigma$ is the standard deviation of the systematic error for a group of machines and $\sigma$ is that of the setup error for a patient population. Here, we need to point out that van Herk et al. did not study the impact of the isocenter shifts on margin determination in their paper. We will take $\Sigma$ as the isocenter shifts' $\Pi_i$ in Eq. (17). The difference between Eq. (17) and Eq. (18) is shown in Fig. 6. In the plot, we take $\overline{W}_{0i} = 0.6\,\mathrm{mm}$. On the one hand, it is apparent that when the residual setup error is very small, our margin is greater than van Herk's. This is due to the fact the mean of the residual setup error is not zero in reality and it was ignored in van Herk's derivation. On the other hand, when the residual setup error is very large, van Herk's margin is greater than ours. This results from the methodology used in van Herk's derivation. His margin formula is for multi-fraction treatment regimens only. Thus, it is not difficult to explain the phenomenon shown in Fig. 6.

The margin formulas developed in Section II. E. have been implemented clinically. The complete procedure involved three steps: (1) measuring the isocenter shift for the linac used; (2) calculating the required margin for the patient; and (3) determining the effect of setup errors on the CTV dose distribution.

## II.H. The procedure to determine frameless SRS margins for a specific machine

Currently, the frameless SRS at our institution is performed with an AKTINA PinPoint Radiosurgery System (AKTINA Medical, Congers, NY). The treatment is delivered on a Trilogy linac (Varian Medical Systems, Palo Alto, CA). The CBCT acquisition protocol is the Varian pre-set CBCT "Pelvis" mode. We have chosen the "Pelvis" mode over the "High Quality Head" mode because the "Pelvis" mode usually produces images with better quality.[9-10] An AlignRT3D



optical surface imaging system (Vision RT, UK) is routinely used for both patient pre-setup and residual setup error measurement. The residual setup errors are defined as the differences between the observed AlignRT values before and after CBCT registrations.

### II.H.1. Measurement of isocenter shifts for a specific machine

For the specific Trilogy machine used, we measured the isocenter shifts between the CBCT and the linac on a weekly basis for three consecutive months. A metal sphere with a 6-mm diameter was positioned near the radiation isocenter using the linac crosshair as guidance. In the first part of the test, a series of Winston-Lutz tests were performed using a $2 \times 2$ cm$^2$ field at the gantry angles of 0°, 90°, 180°, and 270°. The radiation fields were defined by a multi-leaf collimator. The radiation isocenter was assumed to be the intersection point of the central axes of each field. We took the average apparent positions of the sphere to compute an average radiation isocenter position. Knowing the average radiation isocenter position, we imaged the sphere using CBCT at the highest resolution with a slice spacing of 1 mm, a matrix size of $512 \times 512$, and a field-of-view (FOV) of 26 cm. The difference between the two isocenter positions reported by the imaging system and computed with the Winston-Lutz tests was defined as the systematic error. The Winston-Lutz test is one of the standard QA procedures performed at our institution.[27] It was found that all isocenter shifts fluctuated within a very narrow range and were different for different machines, but all were less than 1 mm. Fig. 7 shows the isocenter shifts as a function of time. Similar observations have also been reported in the literature.[28]

### II.H.2. Margin determination for a specific machine



340     In the derivation above, we only considered the isocenter shifts for a normal couch

position (0°). However, for a realistic SRS treatment, non-coplanar beams are often employed in

an effort to achieve a high degree of isodose conformality. Thus, possible couch shifts also exist.

Under this circumstance, the systematic error $\vec{V}_{s0}$ is a function of couch angles. To simplify the

derivation, one can use the maximum components of $\left| \vec{V} \right|$ in our calculation. Here,

345     $\vec{V}_{s0} = \vec{V}_{iso\_shift} + \vec{V}_{couch\_shift}$, containing both isocenter shifts and couch shifts. In our clinical

investigations, we have found that the maximum isocenter shift was approximately 0.5 mm in

each direction and the maximum couch shift was also approximately 0.5 mm. Our method was

similar to the one given in Ref. [6]. However, our measurement results were somewhat different

from theirs. For the frameless SRS cases that we have studied, the collimator angle was fixed.

350     Therefore, we have assumed 1 mm as its systematic error, the worst case scenario in our study.

Because of the technical limitations of current verification systems, it is difficult to

estimate the CBCT-induced residual setup error without a significant amount of extra effort. We,

instead, have used surface imaging for this purpose. The validity of this approach is based on our

clinical observations. We have found that a carefully calibrated surface imaging system can

355     provide not only more accurate translational shifts (in the order of 0.1 mm) than CBCT, but also

consistent and reproducible results. The surface imaging system has three significant advantages

over CBCT. It uses light as the imaging source, employs more than 30,000 points for image

registration, and is real-time in image reconstruction and registration. Thus, it can be repeatedly

applied to the patient, making the residual setup error assessment safe, accurate, and efficient. In

360     this study, the patient's shifts detected by surface imaging measurement after the final CBCT

scan is considered as the residual setup error of the image-guiding system.



The residual setup errors were measured for 30 patients treated at one of our Regional Centers. The results and the required corresponding margins are listed in Table I. As expected, the mean of the residual setup errors was not zero. The maximum mean was about 0.2 mm, leading to a maximum value of $W_{0i} = 1.2$ mm. Substituting this $W_{0i}$ and $\sigma_i$ into Eq. (13) yielded the margin in each direction. For the sake of simplicity, we did not use unequal margins for different directions, a practice similar to the conventional frame-based SRS procedure.[2] Nevertheless, we can use unequal margins in the future if clinically necessary. In the following section, a maximum margin of $C_i \approx 3$ mm is used in our demonstration.

## III. RESULTS AND DISCUSSION

As we mentioned in Section II, several assumptions have been made in our margin formula derivations. Although we argued that these assumptions had been widely used by other investigators in their margin studies, we have nevertheless validated these assumptions by calculating the dose distribution for a previously treated patient. The dose calculation platform used is BrainLab iPlan (BrainLab AG, Germany). The CTV dose coverage was calculated as a function of the isocenter shifts between the CBCT and the planning CT. The cube in Fig. 8 shows the CTV $V_{80\%}$ for a hypothetical situation with a 3-mm setup error in all three directions simultaneously. It represents the worst clinical scenario that occurs with a very small probability in reality. It was intended for demonstration only. In each direction, the probability of the setup error $\geq 3$ mm is: 1-0.98=0.02. With a simple calculation, one can find that the probability for cases in which the setup errors are more than 3 mm in all directions simultaneously is less than 8



$\times\,10^{-6}$. For the eight points on the corners of the cube in Fig. 8, it was found that $V_{80\%}$ was higher

than 95%. Thus, we can conclude that as long as the setup errors fall within this cube, $V_{80\%} \geq$

385    95%.

There are many possible causes for the results shown in Fig. 8. For example, the

imperfect volume expansion from the CTV to the PTV by the treatment planning system may

cause a small fraction of voxels to receive a dose less than the prescribed one even if the setup

error is less than 3 mm in each direction. In addition, the uncertainties in the dose calculation

390    algorithm can also lead to an appreciable dosimetric error for certain types of tissue. We will not

discuss those factors in this paper since they are beyond the scope of our present study.

In our derivation, we assumed that as long as the isocenter shifts were within the

calculated margin (3 mm with a probability of occurrence = 95%), the CTV would receive the

prescribed dose. Thus, the results shown in Fig. 8 may reveal that these assumptions are,

395    perhaps, not realistic and perfect. From a theoretical point of view, it would be valuable to be

able to validate those assumptions based on a group of patients. To this end, we have performed

the same analysis for nine previously treated IG-SRS patients with a total of 11 PTVs (two

patients had two tumors each). Fig. 9 shows the probability distribution of $V_{80\%}$ (3-mm shifts

from the isocenter in all three directions at the same time) for this group of patients. As clearly

400    demonstrated, the majority of the points (71 out of 88) fell in the range of $V_{80\%} \geq 95\%$ with a

probability $\geq 80\%$. Nevertheless, a few points (2 out of 88) did exhibit undesirable CTV

coverage ($V_{80\%} \leq 80\%$). This occurred with a probability $\leq 3\%$ according to Fig. 8. This tells us

that the assumptions have failed for these points. In fact, mathematically, the probability for each

point to have a 3-mm shift in all three directions simultaneously is less than $8 \times 10^{-6}$. In Fig. 10,



405     the average values for $V_{80\%}$ for those 11 PTVs (the average of eight points for each PTV) are

given. Though almost all the average values were larger than 95%, one outlier did fall

somewhere between 70% and 80%. Further investigation has revealed that the CTV of this case

was located at the far posterior side of the brain, where the surface curvature changed very

rapidly, thus violating our third assumption. We should emphasize again that the probability for

410     each point to have a 3-mm shift in all directions simultaneously is extremely small. For this or

similar type of cases, a larger margin is required to compensate for this effect. To further

investigate the phenomenon unveiled in Fig. 10, we have also calculated the additional margins

needed for each of the 11 PTVs. For each tumor, we expanded the CTV-PTV margin such that

our dosimetric objective of $V_{80\%} \geq 99.9\%$ could be easily achieved even if the isocenters were

415     displaced by 3 mm in all three directions simultaneously. The difference between this new

margin and the 3-mm margin was specifically called the additional margin in this paper. For the

11 tumors, there were a total of 88 cases that had 3-mm isocenter shifts in all three directions

simultaneously, giving 88 additional margins. The frequency distribution of these additional

margins is shown in Fig. 11. A Gaussian function was used to fit the additional margin

420     distribution. It was found that the average additional margin needed to achieve our goal was 0.65

mm, with a standard deviation of 0.3 mm. The largest margin corresponded to the case where the

tumor was in the far posterior side of the brain. This approach provided a practical solution to

clinical implementation. To obtain the necessary margins for each specific CTV, one can first

use the method demonstrated in Fig. 8 to estimate the projected $V_{80\%}$ for the corresponding

425     margins. With additional margins, one would be able to achieve $V_{80\%} \geq 99.9\%$. In this study, the

statistical analysis toolbox ANOVA1 in MATLAB (The MathWorks, Inc., MA) was used to test



the statistical significance of the difference between $V_{80\%}$ before and after the addition of this extra margin. The difference with $p < 0.05$ was considered statistically significant. It was found that the addition of extra margins increased the CTV $V_{80\%}$ significantly with $p < 0.01$. However, as previously stated, the setup error of ±3 mm occurring simultaneously in all directions is a very small probability event under normal clinical conditions. Therefore, for practical purpose, we can simply ignore this extra 0.65-mm margin.

       In this paper, we have quantitatively explored the effects of the isocenter shifts between the CBCT and the linac on the margin determination. Because of the potential overlap between the isocenter shifts and setup error and the isocenter shifts being constant over a long period of time, our margin formulas differ from the previous ones, which assumed a Gaussian function for the systematic errors. In our derivation, certain types of the systematic errors were not considered in our mathematical models. These included the target delineation error and the image registration errors. The target delineation error was intentionally excluded in the present study because of the scarcity of clinical data. However, if we assumed that the target delineation error was a Gaussian function, then we could include it in our margin formulas by changing the standard deviation σ to a square root of the sum of the variance of the setup error and the variance of the target delineation error and bring it into our formulas in this paper. Similarly, we could also include the image registration error in our modeling. The validity of this approach, of course, depends on the specific function format of each systematic error, which, in our opinion, should be studied extensively. However, as long as the function format is known, the procedure presented in this paper can be used to drive a new margin formula.



Previous publications on margin formulas concentrated on a group of patients and a group of machines only. van Herk's formula is one of those and has been widely cited in

450     publications. In van Herk's derivation, it was assumed that the contributions from the setup error and the systematic error were different. Our margin formula aims at a single-fraction IG-SRS and a specific machine. In our derivation, it was assumed that the contribution from the setup error and the systematic error were the same. Therefore, it is not surprising that there are differences between our formulas and van Herk's. In a previous publication, it was assumed that

455     the nonzero constant systematic error could be corrected by couch shifting. However, this technique is not effective for the sub-millimeter systematic errors (or residual systematic error) because current 4-DOF couches are not equipped with high-precision stepping motors. Nevertheless, this systematic error is explicitly included in our margin formulas. On the other hand, this non-zero constant systematic error is ignored in the previous derivations.[11–21]

460     People are cautioned if they intend to apply our margin formulas to their IG-SRS programs and machines. They must determine their specific systematic error and verify its function format. For example, for the CBCT setup technique, they would need to determine their own isocenter shifts and make sure that these shifts are almost constant. Subsequently, they can substitute their isocenter shifts into the formulas given in this paper. For multiple systematic

465     errors, the best way to include them in the margin formulas is to determine the function format of the combined systematic error. If the combined systematic error is a constant, then it can be used to replace the isocenter shifts in the formulas in this paper.

As the treatment margin is a statistical quantity, it is, therefore, aimed at benefiting the majority of the patient population, rather than a specific one. In this study, it is assumed that the



470   setup errors follow a Gaussian distribution. The margin formula is derived such that it will cover

setup errors in up to 95% of all SRS patients. Consequently, by definition, those 95% patients

will be over-compensated or overdosed. In fact, only those patients whose setup errors are

exactly the same as the expansion margin will not be overcompensated, nor will these tumors be

missed. Overcompensation is an intrinsic phenomenon or deficiency in margin formulas

475   Therefore, we strongly urge our readers and colleagues to carefully examine their patients' setup

errors in order to arrive at their own specific expansion margins.  However, for multi-

fractionated treatment protocols, there is indeed one practical approach to avoid or alleviate

margin overcompensation. That is the patient specific margin. In this context, one could derive a

margin based on $n$ treated fractions for that particular patient and then use this derived value for

480   adaptive treatment planning.[29] Nevertheless, this method is not applicable to single fractionated

treatment.

## IV. CONCLUSIONS

The margin definition described in this paper is machine-specific and more appropriate

485   for single-fraction IG-SRS. Two different types of volume expansion strategies have been

presented in the paper: (1) asymmetric expansion and (2) symmetric expansion that includes the

nonzero constant systematic error. Margin formulas for single-fraction frameless IG-SRS and a

group of machines have also been derived. It has been found that this nonzero constant machine

systematic error made the margin formulas more complex than the previous ones. Our margin

490   formulas are innovative and have never been reported previously. Our methodology eliminates

the assumption used in the previous margin formula derivations, i.e., the mean systematic error



(or mean residual systematic error) is zero, thereby, making it more general and practical for clinical applications.

Table I. The residual setup errors measured with an AlignRT system for patients treated at one of our Regional Centers and their corresponding margins calculated with our method. Unit: mm

| Residual setup error | LR (left-Right) | SI (superior-inferior) | AP (anterior-posterior) |
|---|---|---|---|
| Mean | -0.17 | 0.10 | -0.03 |
| Stand Deviation σ | 0.59 | 0.81 | 0.43 |
| Margin | 2.41 | 2.87 | 2.08 |

580



## Figure Legends

### Figure 1

Examples of 3D symmetric expansion, 1D symmetric expansion, and 1D asymmetric expansion.

585    The 3D expansion is a symmetric expansion of the base volume by the same distance in all

directions (left), whereas 1D expansion can be symmetric (middle) or asymmetric (right). The

inner circle is the CTV, and the outside contour is the PTV. For demonstration purposes, only 2D

images are shown.

590    ### Figure 2

Behavior of the margin parameter $b_1(W_{0i})$ as a function of $W_{0i}$.

### Figure 3

Behavior of the margin parameter $b_2(W_{0i})$ as a function of $W_{0i}$.

595

### Figure 4

The exact and approximated solutions computed by Eq. (7) (circle) and Eq. (13) (solid line),

respectively. The differences between those two solutions are indicated by the dotted line.

600    ### Figure 5

A demonstration of the possible dose error occurrence when using the "CTV copy" technique.

The left image is the planning CT with the original CTV. The beam arrangement should cover



the whole CTV. The CTV is moved to a new location because of the isocenter shift between the CBCT and the linac (right). A and B indicate the beam edges.

605

**Figure 6**

The differences between van Herk's formula and our formula for a group machine and a group patient as a function of $\sigma$. But different methodology has been used in the derivation of van Herk (Eq. (18)) and ours (Eq. (17)).

610

**Figure 7**

The measured isocenter shifts between the CBCT and the linac as a function of time. Isocenter shifts remain approximately constant over a three-month period.

615     **Figure 8**

The $V_{80\%}$ values when the linac and CBCT isocenters differ by 3 mm in all three directions simultaneously. The center of the cube represents the linac isocenter. Each point on the corner represents the worst clinical scenario that barely occurs in reality (with a probability less than $8 \times 10^{-6}$).

620

**Figure 9**

Frequency of the CTV $V_{80\%}$ of 11 clinical tumors for the extreme case when there were 3-mm shifts in all directions simultaneously. A total of 88 points were obtained for these 11 clinical tumors.



625

**Figure 10**

The average CTV $V_{80\%}$ for each of the 11 tumors in nine patients when the setup errors were 3

mm in all three directions simultaneously. Each point represents the worst clinical scenario.

630     **Figure 11**

The frequency distribution of the additional margins.  A Gaussian function was used to fit the

distribution.

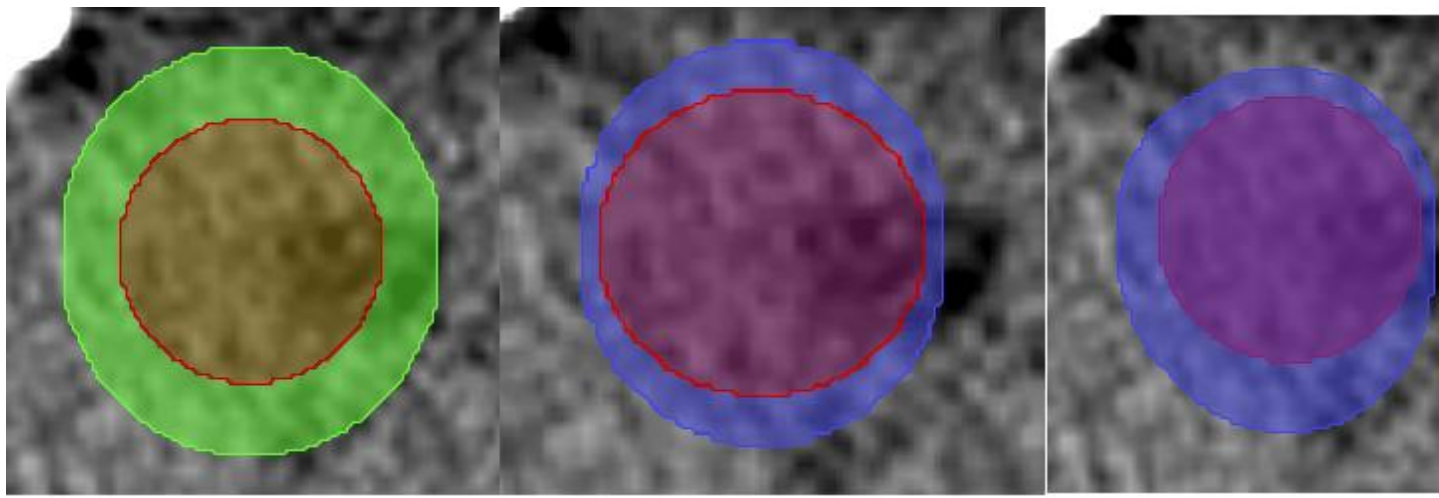

635

Fig.1



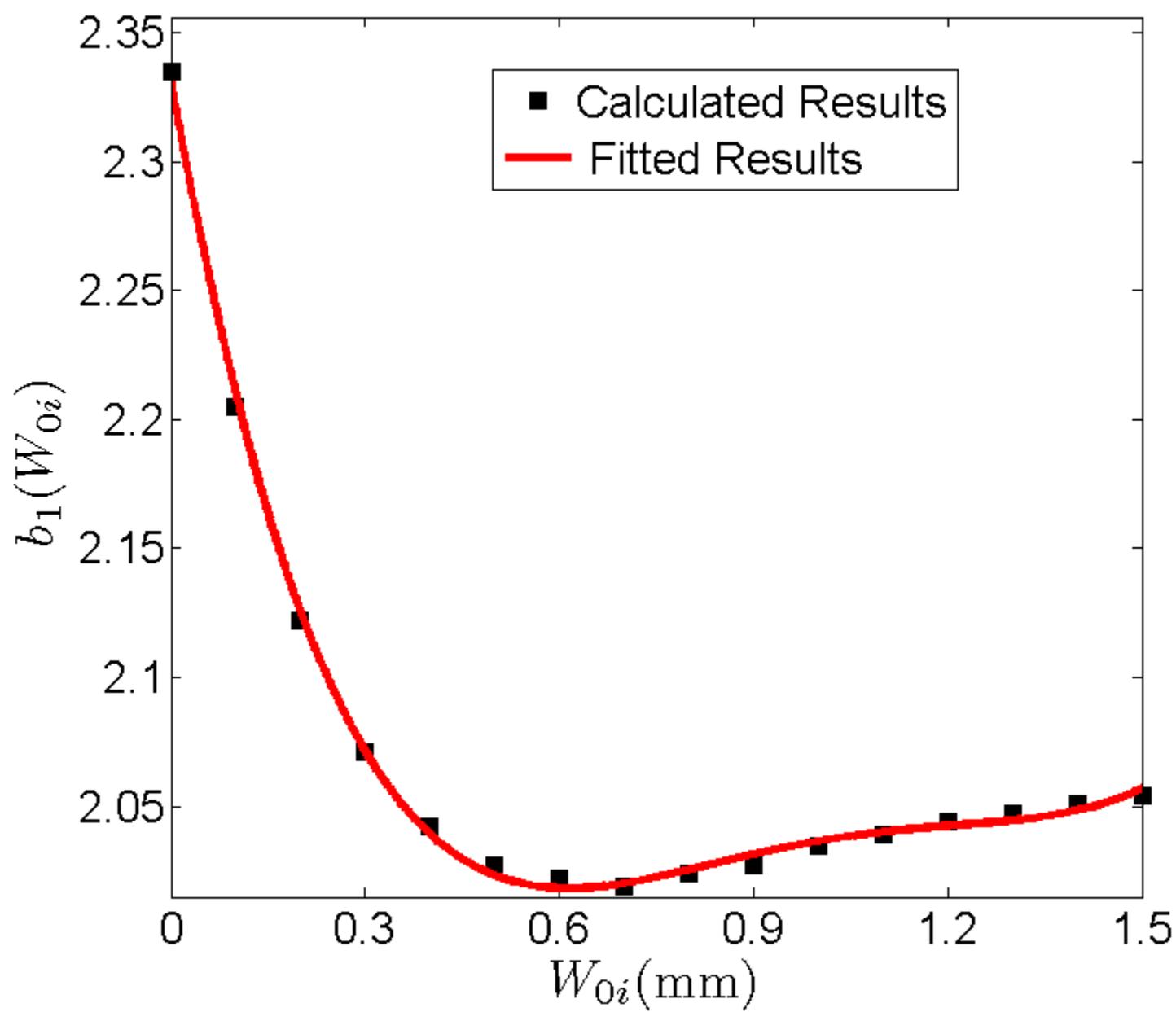

Fig. 2





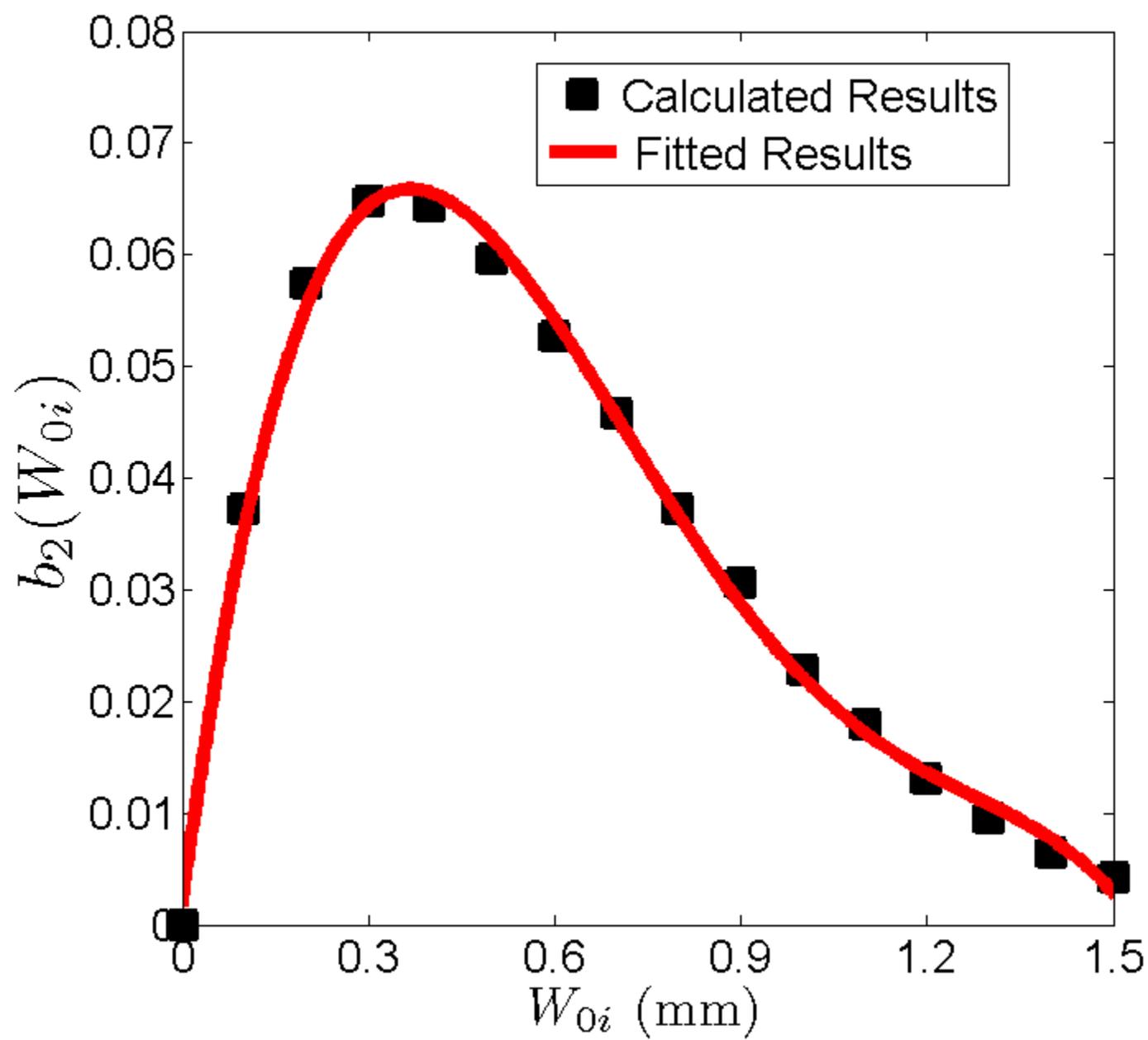

Fig. 3



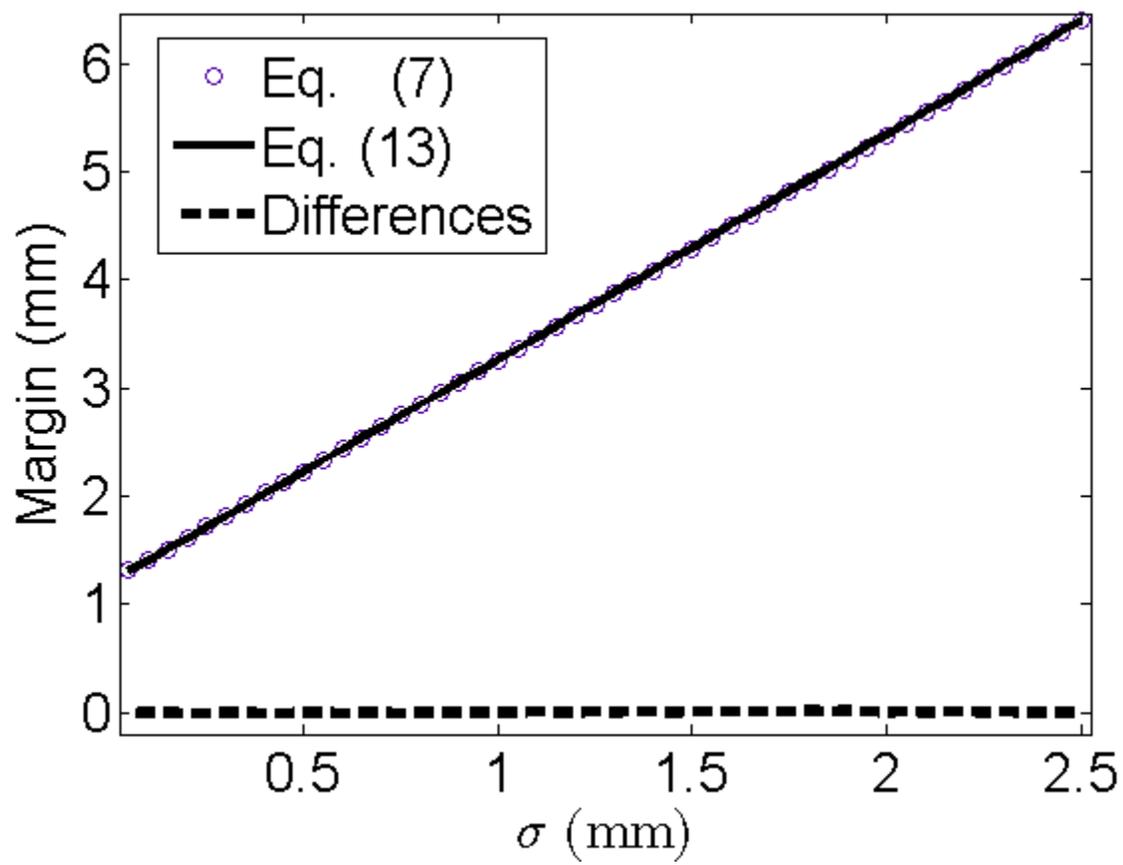







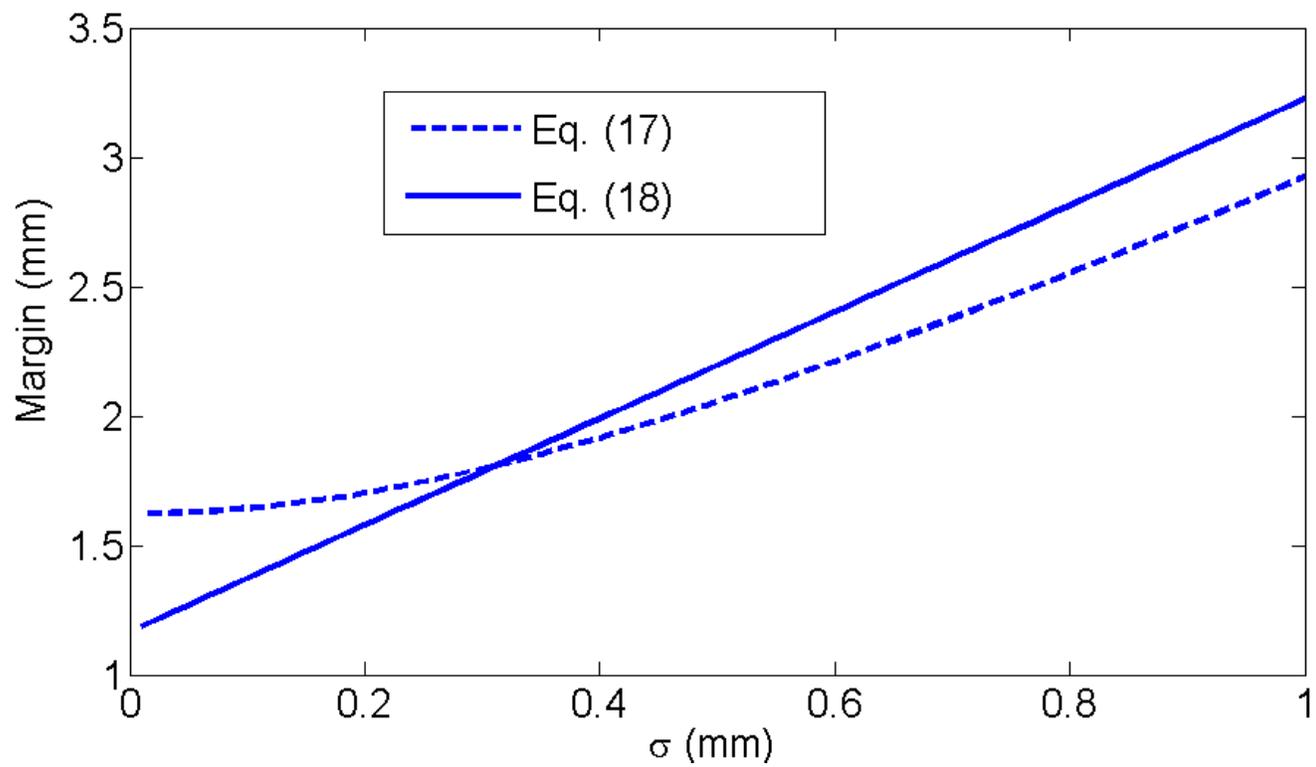

Fig. 5

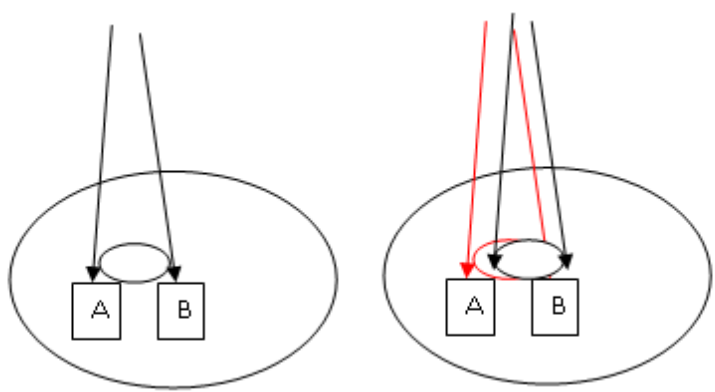

655    Fig.6



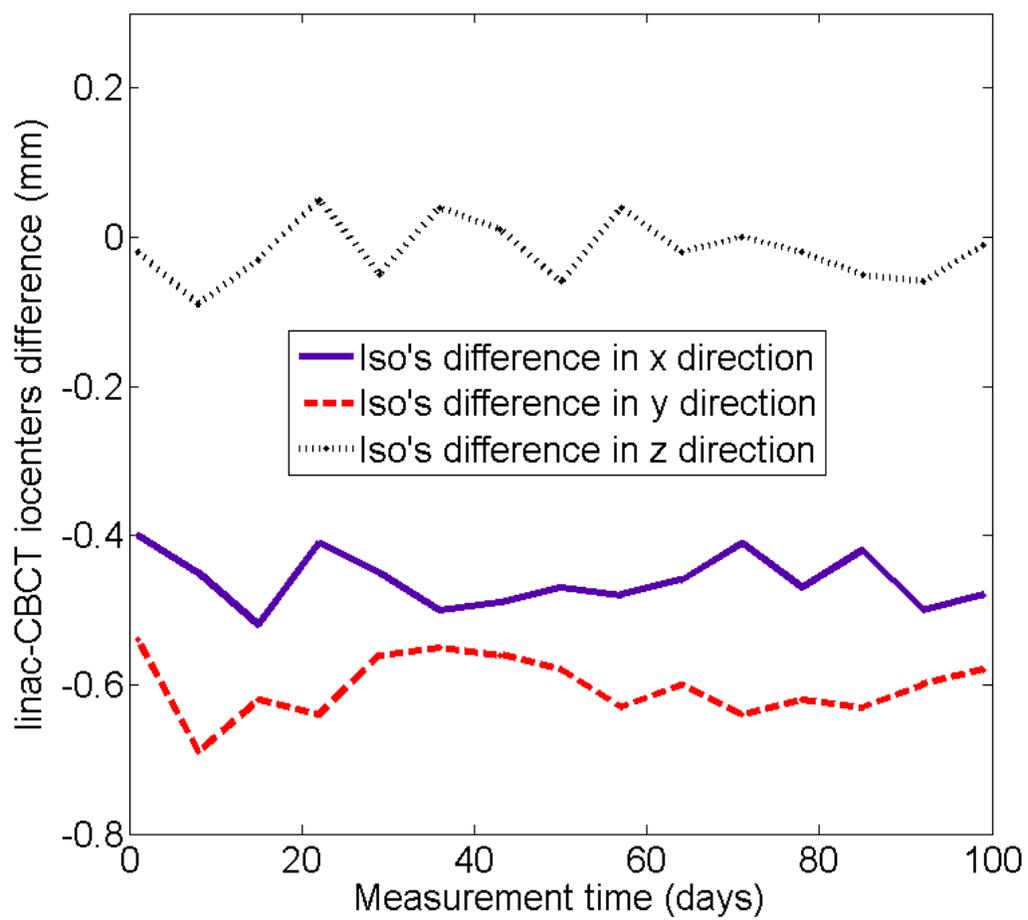

Fig. 7



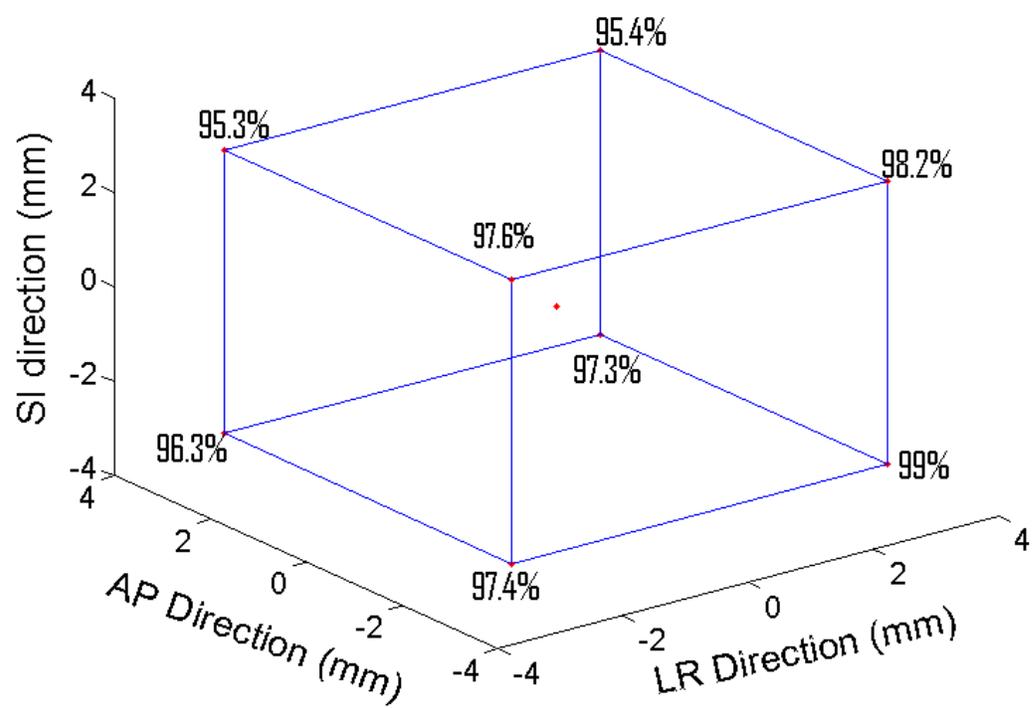



Fig. 8



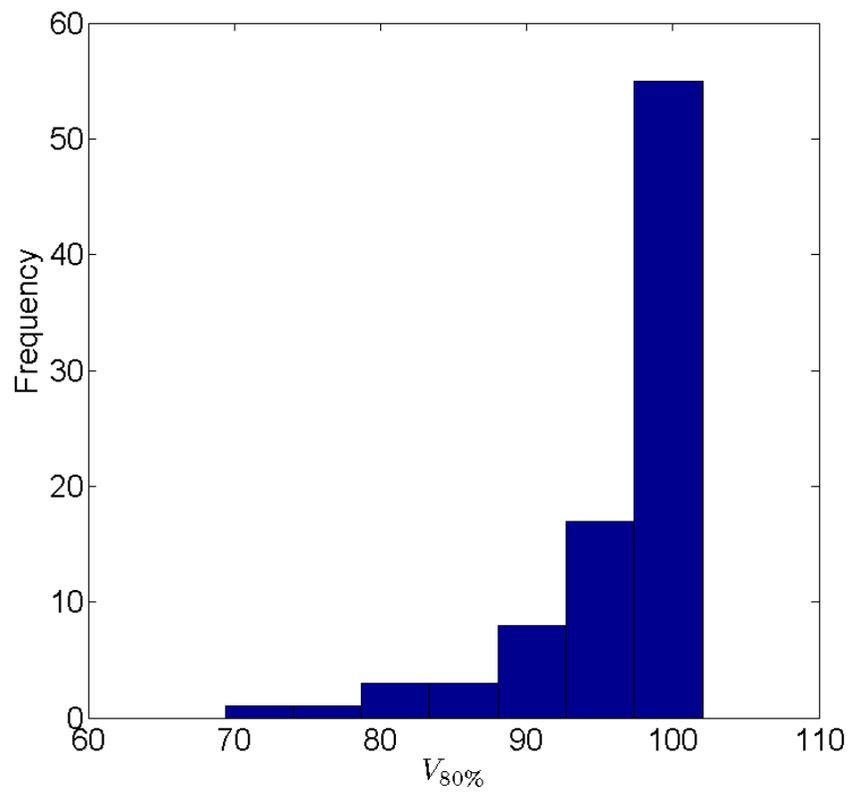

Fig. 9



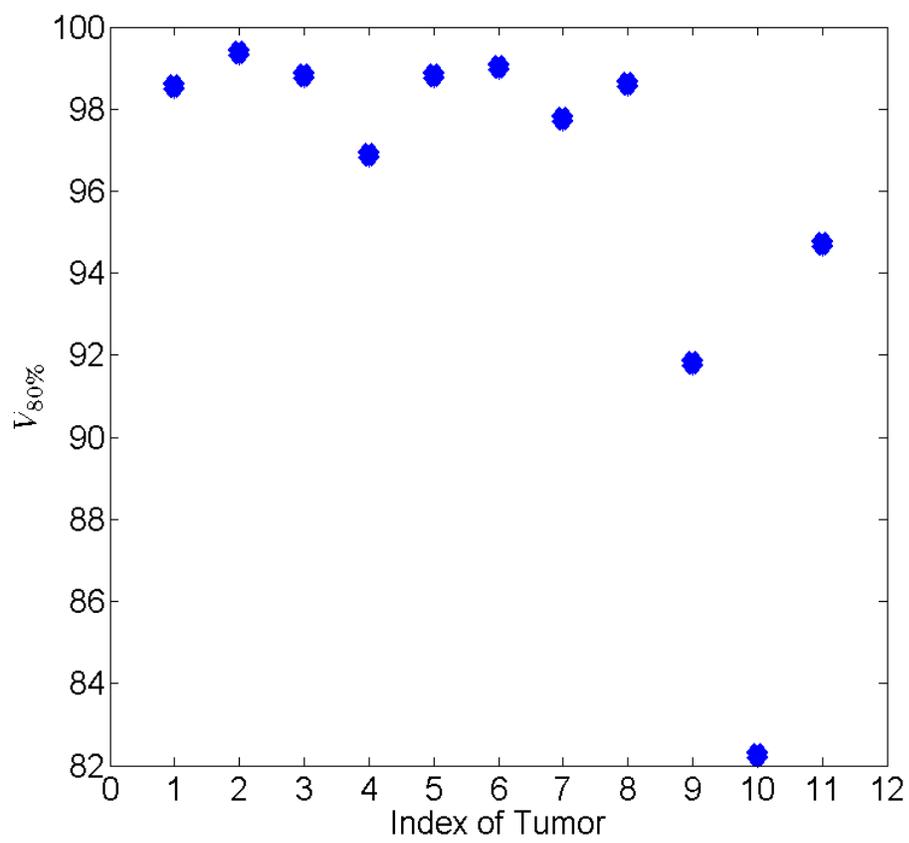

    Fig. 10







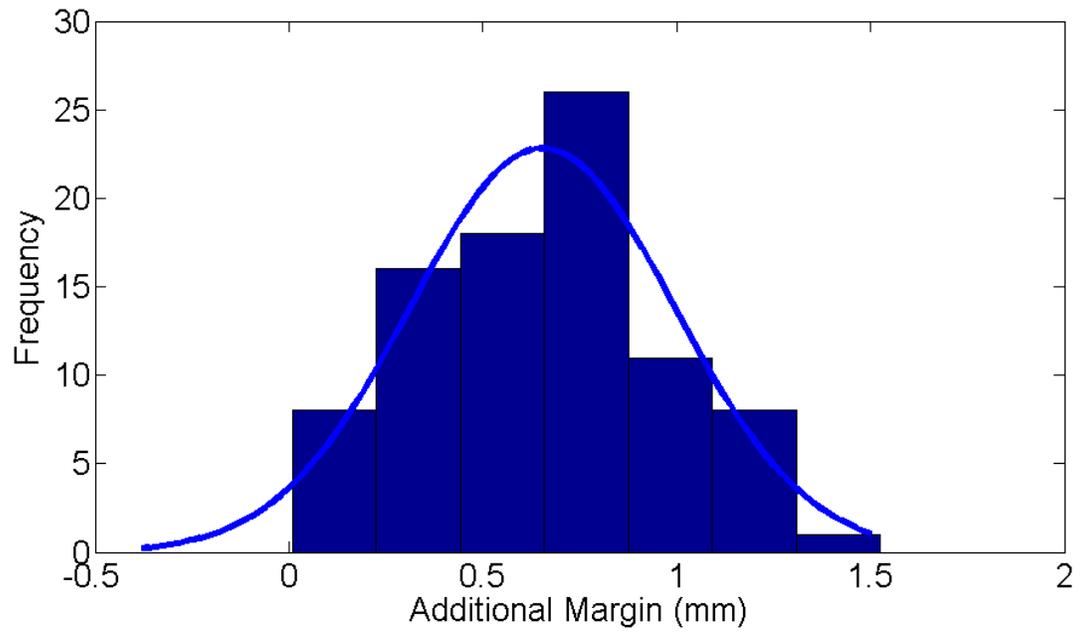

Fig. 11